\documentclass[aps,prl,twocolumn,tightenlines,superscriptaddress,showpacs,byrevtex]{revtex4}

\usepackage{rotating}
\usepackage{longtable}
\usepackage{amsmath}
\usepackage{graphicx} 
\usepackage{dcolumn}  
\usepackage{multirow}
\usepackage{color}

\newcommand{\mev}{\mathrm{MeV}}

\newcommand{\mevm}{\mathrm{MeV}/c^2}
\newcommand{\gev}{\mathrm{GeV}}

\newcommand{\gevm}{\mathrm{GeV}/c^2}

\newcommand{\ee}{e^+e^-}
\newcommand{\uu}{\mu^+\mu^-}
\newcommand{\pp}{\pi^+\pi^-}
\newcommand{\U}{\Upsilon}
\newcommand{\Uf}{\Upsilon(5S)}
\newcommand{\Uo}{\Upsilon(1S)}
\newcommand{\Un}{\Upsilon(nS)}
\newcommand{\Ut}{\Upsilon(2S)}
\newcommand{\Uth}{\Upsilon(3S)}

\newcommand{\mmpp}{M_{\rm miss}(\pi^+\pi^-)}
\newcommand{\mmp}{M_{\rm miss}(\pi)}
\newcommand{\mmpip}{M_{\rm miss}(\pi^+)}
\newcommand{\mmpim}{M_{\rm miss}(\pi^-)}

\newcommand{\hb}{h_b(1P)}

\newcommand{\hbp}{h_b(2P)}
\newcommand{\hbn}{h_b(mP)}
\newcommand{\ks}{K^0_S}

\newcommand{\pipm}{\pi^{\pm}}
\newcommand{\pimp}{\pi^{\mp}}

\newcommand{\fb}{\mathrm{fb}^{-1}}
\newcommand{\etal}{\em et al.}

\newcommand{\zb}{Z_b}
\newcommand{\zbo}{Z_b(10610)}
\newcommand{\zbt}{Z_b(10650)}

\begin{document}

\title{\boldmath Observation of two charged bottomonium-like resonances in $\Uf$ decays}

\date{\today}

\begin{abstract}
\noindent
We report the observation of two narrow structures in the mass spectra of the
$\pi^\pm\Un$ ($n=1,2,3$) and $\pi^\pm\hbn$ ($m=1,2$) pairs that are produced
in association with a single charged pion in $\Uf$ decays. The measured masses
and widths of the two structures averaged over the five final states are 
$M_1=(10607.2\pm2.0)\,\mevm$, $\Gamma_1=(18.4\pm2.4)\,\mev$ and 
$M_2=(10652.2\pm1.5)\,\mevm$, $\Gamma_2=(11.5\pm2.2)\,\mev$. 
The results are obtained with a $121.4\,{\rm fb}^{-1}$ data sample collected
with the Belle detector in the vicinity of the $\Uf$ resonance at the KEKB
asymmetric-energy $\ee$ collider.
\end{abstract}

\pacs{14.40.Pq, 13.25.Gv, 12.39.Pn}

\affiliation{Budker Institute of Nuclear Physics SB RAS and Novosibirsk State University, Novosibirsk 630090}
\affiliation{Faculty of Mathematics and Physics, Charles University, Prague}
\affiliation{University of Cincinnati, Cincinnati, Ohio 45221}
\affiliation{Department of Physics, Fu Jen Catholic University, Taipei}
\affiliation{Justus-Liebig-Universit\"at Gie\ss{}en, Gie\ss{}en}
\affiliation{Gifu University, Gifu}
\affiliation{Gyeongsang National University, Chinju}
\affiliation{Hanyang University, Seoul}
\affiliation{University of Hawaii, Honolulu, Hawaii 96822}
\affiliation{High Energy Accelerator Research Organization (KEK), Tsukuba}
\affiliation{Indian Institute of Technology Madras, Madras}
\affiliation{Institute of High Energy Physics, Chinese Academy of Sciences, Beijing}
\affiliation{Institute of High Energy Physics, Vienna}
\affiliation{Institute of High Energy Physics, Protvino}
\affiliation{INFN - Sezione di Torino, Torino}
\affiliation{Institute for Theoretical and Experimental Physics, Moscow}
\affiliation{J. Stefan Institute, Ljubljana}
\affiliation{Kanagawa University, Yokohama}
\affiliation{Institut f\"ur Experimentelle Kernphysik, Karlsruher Institut f\"ur Technologie, Karlsruhe}
\affiliation{Korea Institute of Science and Technology Information, Daejeon}
\affiliation{Korea University, Seoul}
\affiliation{Kyoto University, Kyoto}
\affiliation{Kyungpook National University, Taegu}
\affiliation{\'Ecole Polytechnique F\'ed\'erale de Lausanne (EPFL), Lausanne}
\affiliation{Faculty of Mathematics and Physics, University of Ljubljana, Ljubljana}
\affiliation{Luther College, Decorah, Iowa 52101}
\affiliation{University of Maribor, Maribor}
\affiliation{Max-Planck-Institut f\"ur Physik, M\"unchen}
\affiliation{University of Melbourne, School of Physics, Victoria 3010}
\affiliation{Graduate School of Science, Nagoya University, Nagoya}
\affiliation{Kobayashi-Maskawa Institute, Nagoya University, Nagoya}
\affiliation{Nara Women's University, Nara}
\affiliation{National Central University, Chung-li}
\affiliation{National United University, Miao Li}
\affiliation{Department of Physics, National Taiwan University, Taipei}
\affiliation{H. Niewodniczanski Institute of Nuclear Physics, Krakow}
\affiliation{Nippon Dental University, Niigata}
\affiliation{Niigata University, Niigata}
\affiliation{Osaka City University, Osaka}
\affiliation{Osaka University, Osaka}
\affiliation{Pacific Northwest National Laboratory, Richland, Washington 99352}
\affiliation{Panjab University, Chandigarh}
\affiliation{Research Center for Nuclear Physics, Osaka}
\affiliation{University of Science and Technology of China, Hefei}
\affiliation{Seoul National University, Seoul}
\affiliation{Sungkyunkwan University, Suwon}
\affiliation{School of Physics, University of Sydney, NSW 2006}
\affiliation{Tata Institute of Fundamental Research, Mumbai}
\affiliation{Excellence Cluster Universe, Technische Universit\"at M\"unchen, Garching}
\affiliation{Tohoku Gakuin University, Tagajo}
\affiliation{Tohoku University, Sendai}
\affiliation{Department of Physics, University of Tokyo, Tokyo}
\affiliation{Tokyo Institute of Technology, Tokyo}
\affiliation{Tokyo Metropolitan University, Tokyo}
\affiliation{Tokyo University of Agriculture and Technology, Tokyo}
\affiliation{CNP, Virginia Polytechnic Institute and State University, Blacksburg, Virginia 24061}
\affiliation{Yonsei University, Seoul}
  \author{A.~Bondar}\affiliation{Budker Institute of Nuclear Physics SB RAS and Novosibirsk State University, Novosibirsk 630090} 
  \author{A.~Garmash}\affiliation{Budker Institute of Nuclear Physics SB RAS and Novosibirsk State University, Novosibirsk 630090} 
  \author{R.~Mizuk}\affiliation{Institute for Theoretical and Experimental Physics, Moscow} 
  \author{D.~Santel}\affiliation{University of Cincinnati, Cincinnati, Ohio 45221} 
  \author{K.~Kinoshita}\affiliation{University of Cincinnati, Cincinnati, Ohio 45221} 
  \author{I.~Adachi}\affiliation{High Energy Accelerator Research Organization (KEK), Tsukuba} 
  \author{H.~Aihara}\affiliation{Department of Physics, University of Tokyo, Tokyo} 
  \author{K.~Arinstein}\affiliation{Budker Institute of Nuclear Physics SB RAS and Novosibirsk State University, Novosibirsk 630090} 
  \author{D.~M.~Asner}\affiliation{Pacific Northwest National Laboratory, Richland, Washington 99352} 
  \author{T.~Aushev}\affiliation{Institute for Theoretical and Experimental Physics, Moscow} 
% ...
\collaboration{The Belle Collaboration}

\maketitle

{\renewcommand{\thefootnote}{\fnsymbol{footnote}}}
\setcounter{footnote}{0}

Recent studies of heavy quarkonium have produced a number of surprises
and puzzles~\cite{nora}, including some associated with $\Uf$ decays to
non-$B\overline{B}$ final states. The Belle Collaboration reported the
observation of anomalously high rates for $\Uf\to\Un\pp$ 
($n=1,2,3$)~\cite{Belle_ypipi} and $\Uf\to\hbn\pp$ ($m=1,2$)~\cite{Belle_hb}
transitions. If the $\Un$ signals are attributed entirely to $\Uf$ decays,
the measured partial decay widths $\Gamma[\Uf\to\Un\pp]\sim0.5\,\mev$ are
about two orders of magnitude larger than typical widths for dipion
transitions among the four lower $\U(nS)$ states.
Furthermore, the processes $\Uf\to\hbn\pp$, which require a heavy-quark
spin flip, are found to have rates that are comparable to those for the
heavy-quark spin conserving transitions $\Uf\to\Un\pp$~\cite{Belle_hb}.
These observations differ from {\em apriori} theoretical expectations and
strongly suggest that exotic mechanisms are contributing to $\Uf$ decays.
We report results of resonant substructure studies of $\Uf\to\Un\pp$
($n=1,2,3$) and $\Uf\to\hbn\pp$ ($m=1,2$) decays~\cite{foot1}. We use a 
$121.4\,\fb$ data sample collected on or near the peak of the $\Uf$ 
resonance ($\sqrt{s}\sim 10.865\,\gev$) with the Belle detector at the
KEKB asymmetric energy $\ee$ collider~\cite{KEKB}.

The Belle detector is a large-solid-angle magnetic spectrometer that
consists of a silicon vertex detector, a central drift chamber, an array
of aerogel threshold Cherenkov counters, a barrel-like arrangement of
time-of-flight scintillation counters, and an electromagnetic calorimeter
comprised of CsI(Tl) crystals located inside a superconducting solenoid
that provides a 1.5~T magnetic field. An iron flux-return located outside
the coil is instrumented to detect $K_L^0$ mesons and to identify muons.
The detector is described in detail elsewhere~\cite{BELLE}.

To reconstruct $\Uf\to\Un\pp$, $\Un\to\uu$ candidates we select events
with four charged tracks with zero net charge that are consistent with
coming from the interaction point. Charged pion and muon candidates
are required to be positively identified. Exclusively reconstructed
events are selected by the requirement $|\mmpp-M(\uu)|<0.2$~GeV/$c^2$,
where $\mmpp$ is the missing mass recoiling against the $\pp$ system
calculated as $\mmpp=\sqrt{(E_{\rm c.m.}-E_{\pp}^*)^2-p_{\pp}^{*2}}$,
$E_{\rm c.m.}$ is the center-of-mass (c.m.) energy and $E^*_{\pp}$ and
$p^*_{\pp}$ are the energy and momentum of the $\pp$ system measured
in the c.m.\ frame. Candidate $\Uf\to\Un\pp$ events are selected by
requiring $|\mmpp-m_{\Un}|<0.05\,\gevm$, where $m_{\Un}$ is the mass
of an $\Un$ state~\cite{PDG}. Sideband regions are defined as
$0.05\,\gevm<|\mmpp-m_{\Un}|<0.10\,\gevm$.  To remove background due
to photon conversions in the innermost parts of the Belle detector we
require $M^2(\pp)>0.20/0.14/0.10\,\gevm$ for a final state with an
$\Uo, \Ut, \Uth$, respectively.

Amplitude analyses of the three-body $\Uf\to\Un\pp$ decays reported here are
performed by means of unbinned maximum likelihood fits to two-dimensional 
$M^2[\Un\pi^+]$ vs.\ $M^2[\Un\pi^-]$ Dalitz distributions. The fractions of
signal events in the signal region are determined from fits to the 
corresponding $\mmpp$ spectrum and are found to be $0.937\pm0.015$(stat.),
$0.940\pm0.007$(stat.), $0.918\pm0.010$(stat.) for final states with $\Uo$,
$\Ut$, $\Uth$, respectively. The variation of reconstruction efficiency
across the Dalitz plot is determined from a GEANT-based MC 
simulation~\cite{GEANT}. The distribution of background events is
determined using events from the $\Un$ sidebands and found to be uniform
(after efficiency correction) across the Dalitz plot.

\begin{figure}[!t]
  \centering
\hspace*{-1mm}
  \includegraphics[width=0.23\textwidth]{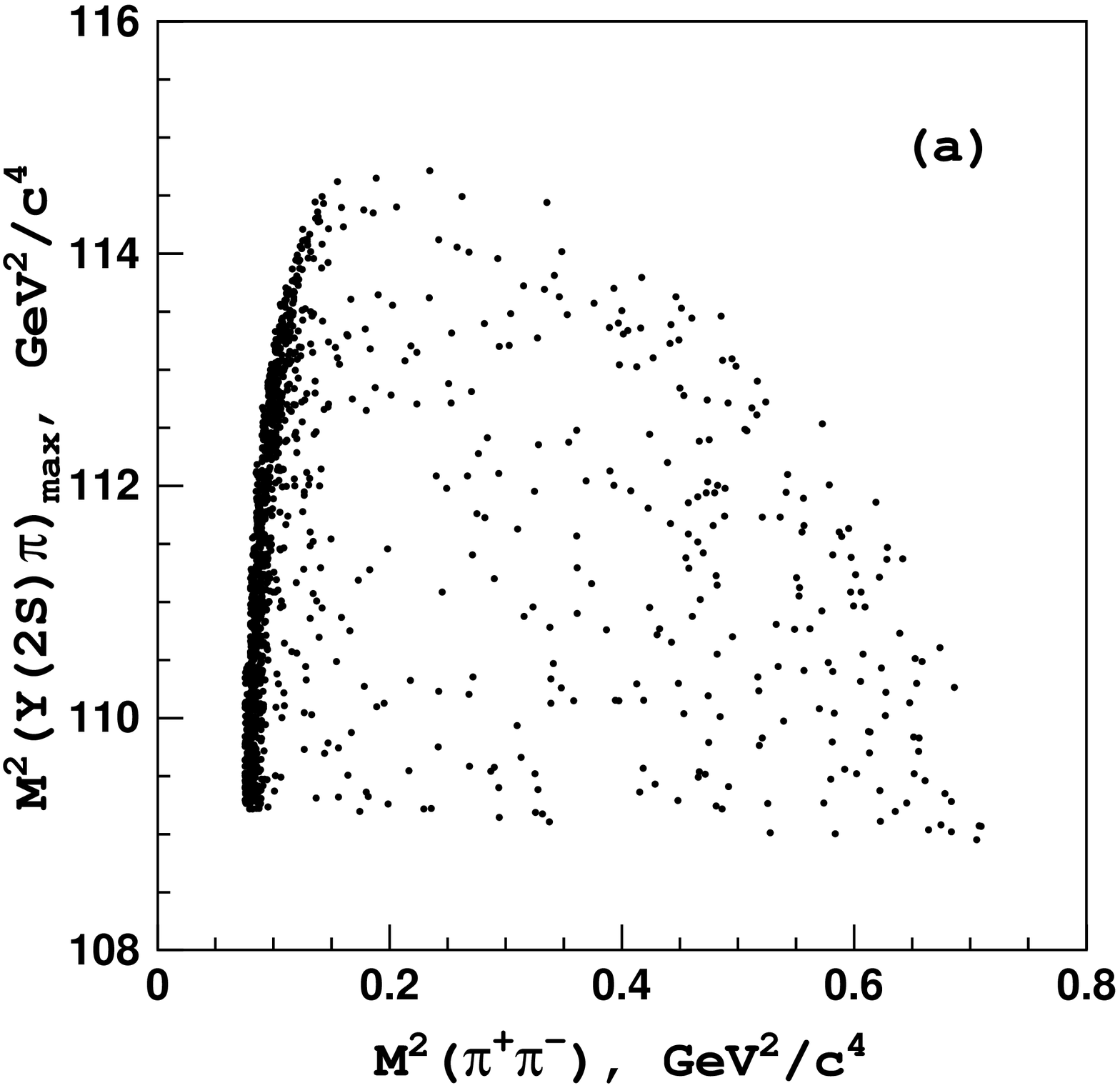} \hfill
  \includegraphics[width=0.23\textwidth]{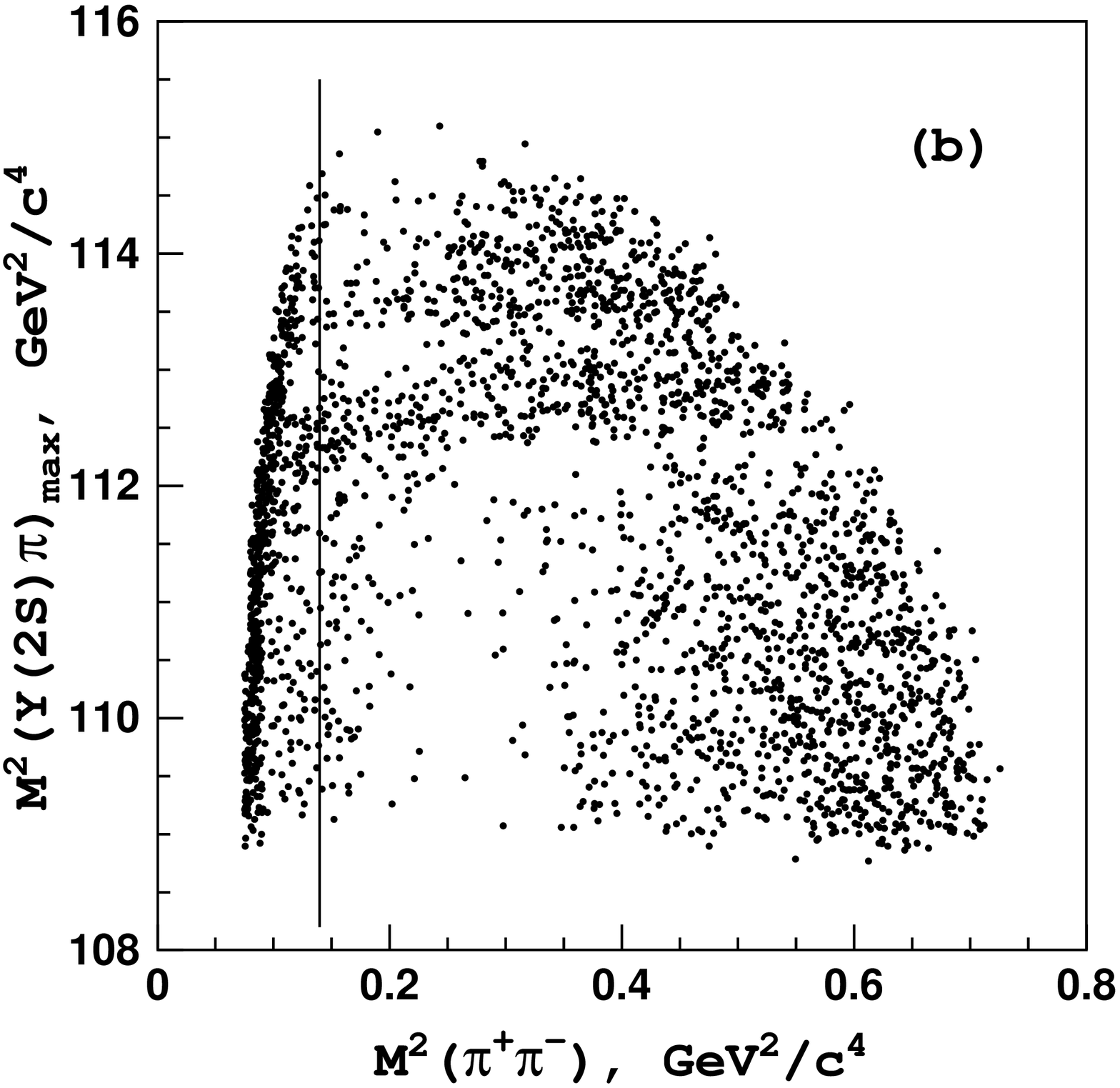}
  \caption{Dalitz plots for $\Ut\pp$ events in the (a) $\Ut$ sidebands; (b)
  $\Ut$ signal region. Events to the left of the vertical line are excluded.}
\label{fig:ynspp-b-dp}
\end{figure}

\begin{figure}[!t]
  \centering
  \includegraphics[width=0.23\textwidth]{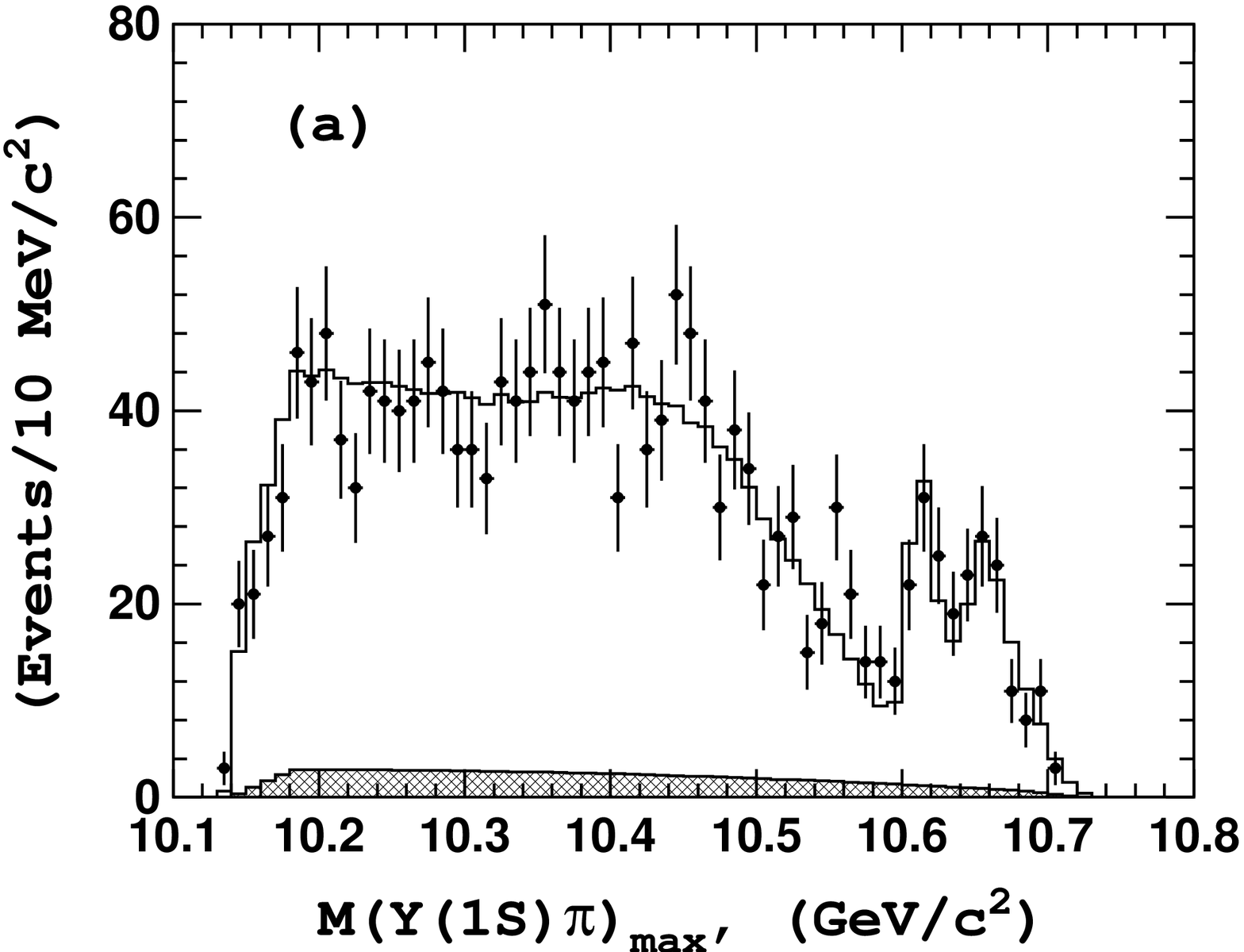} \hfill
  \includegraphics[width=0.23\textwidth]{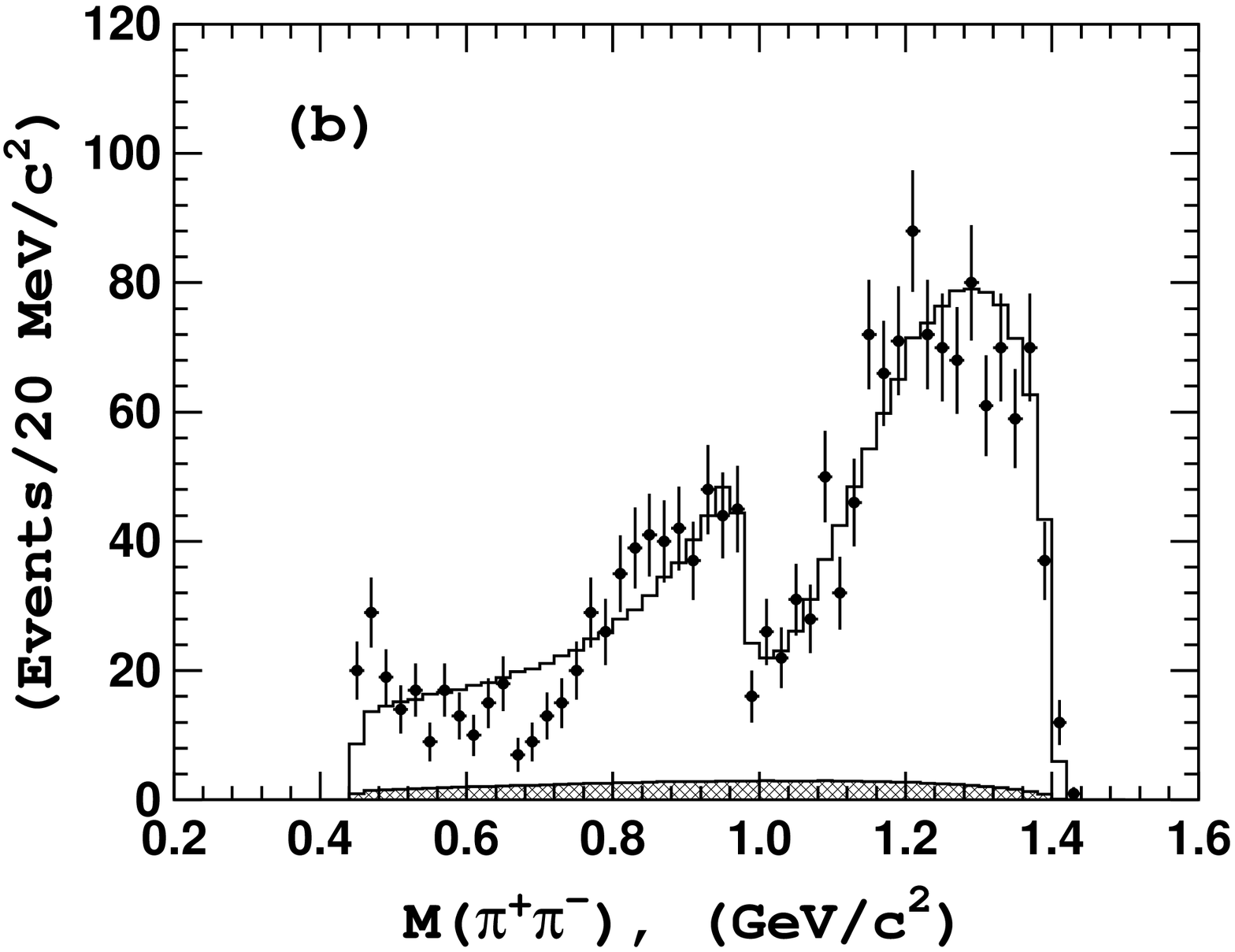} \\
  \includegraphics[width=0.23\textwidth]{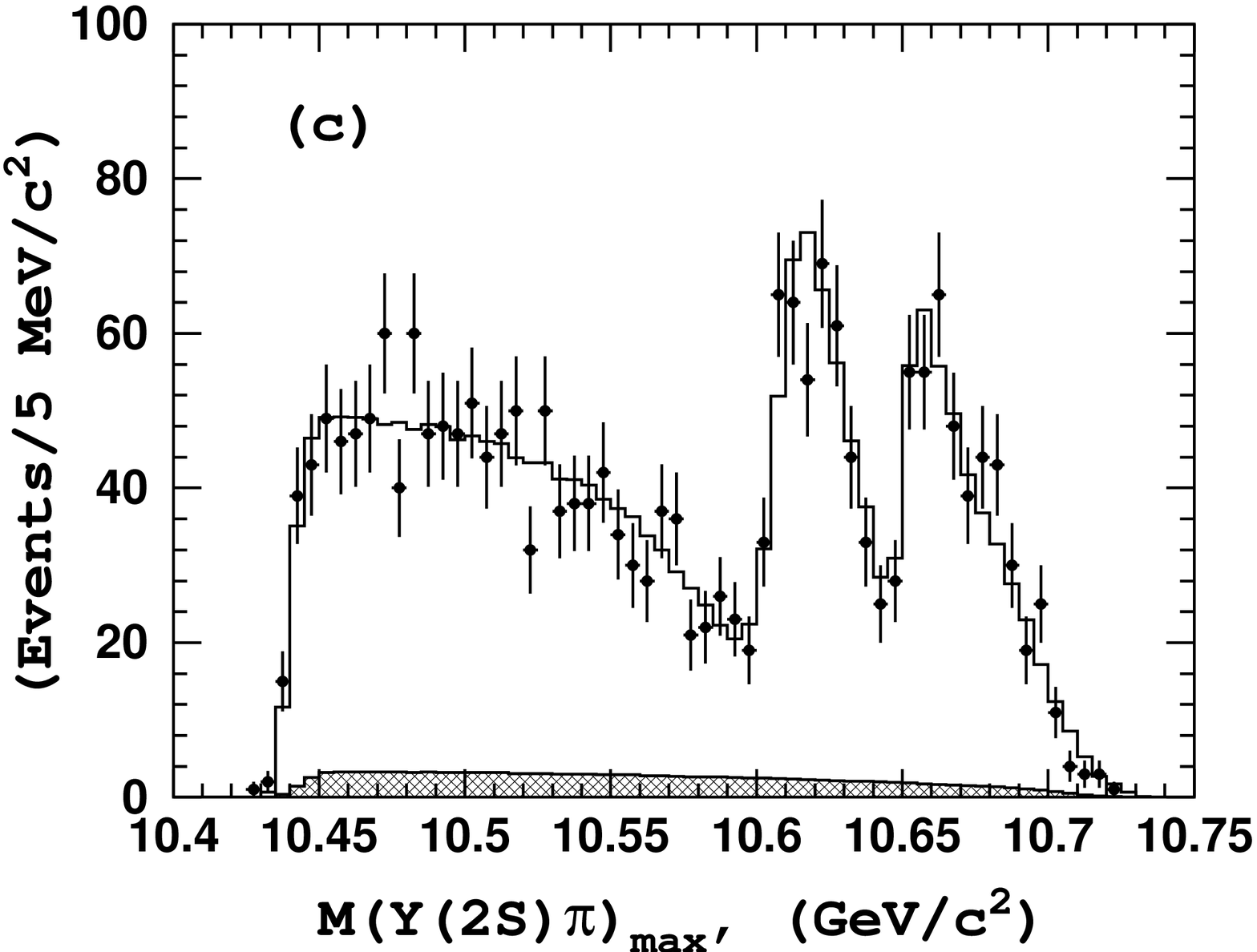} \hfill
  \includegraphics[width=0.23\textwidth]{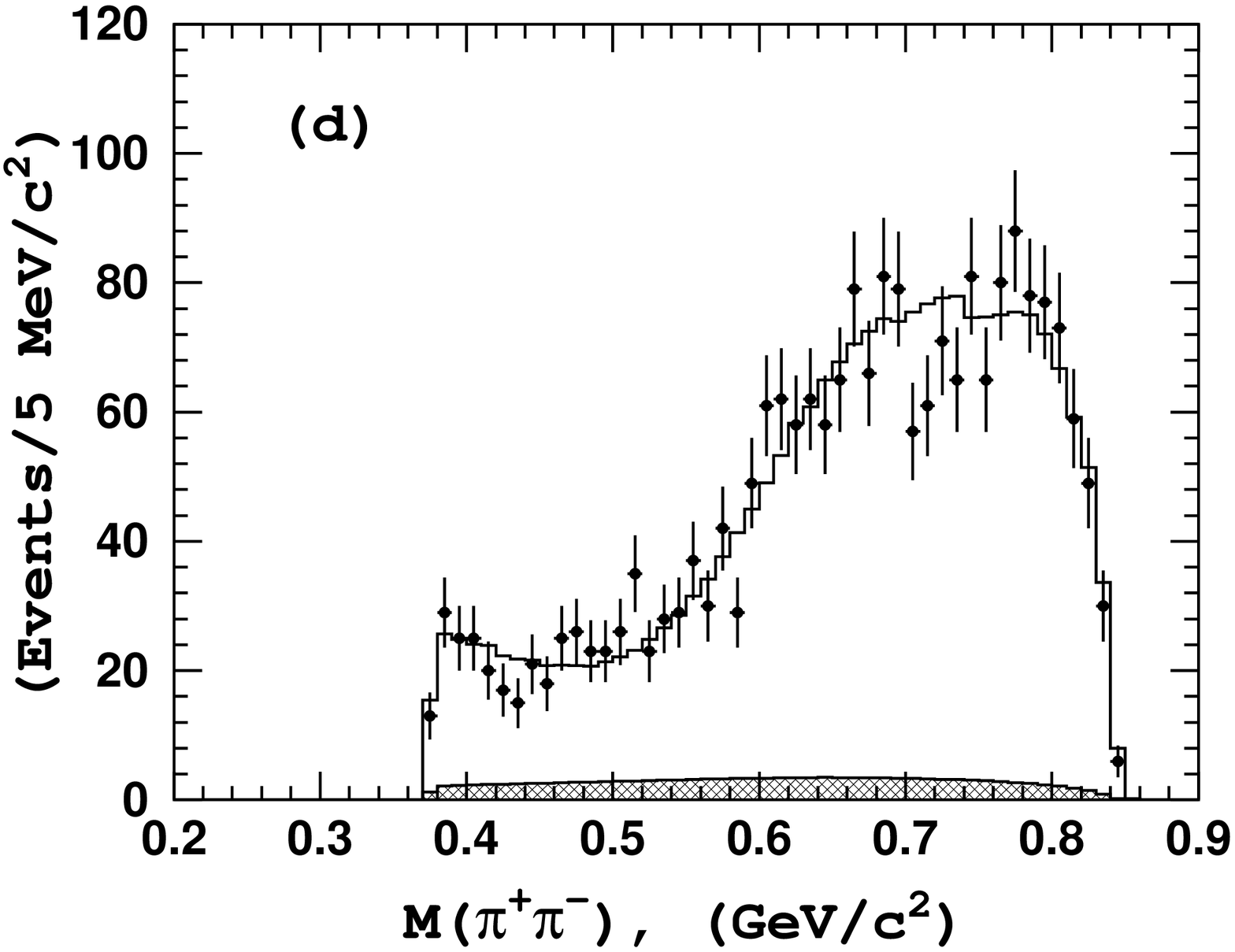} \\
  \includegraphics[width=0.23\textwidth]{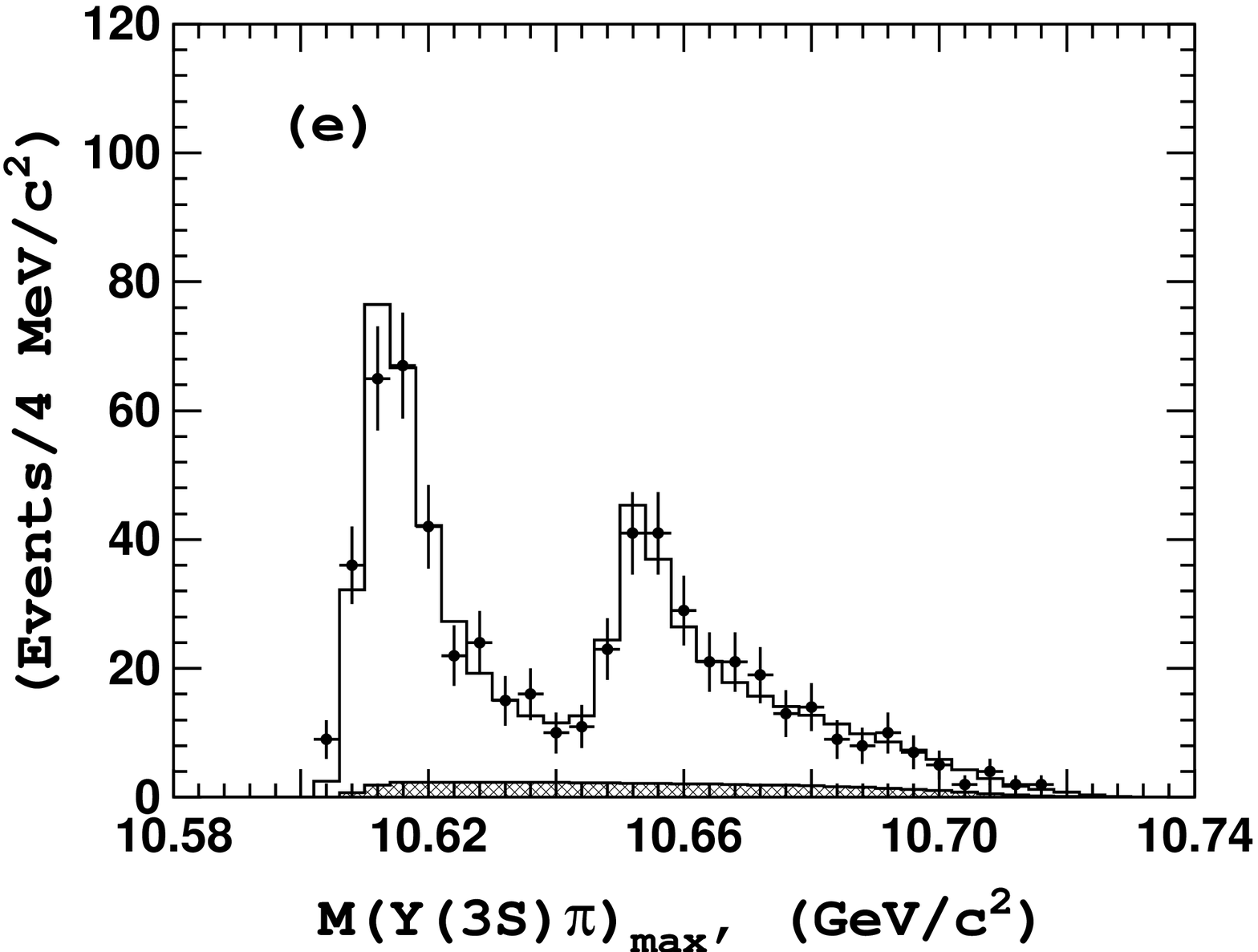} \hfill
  \includegraphics[width=0.23\textwidth]{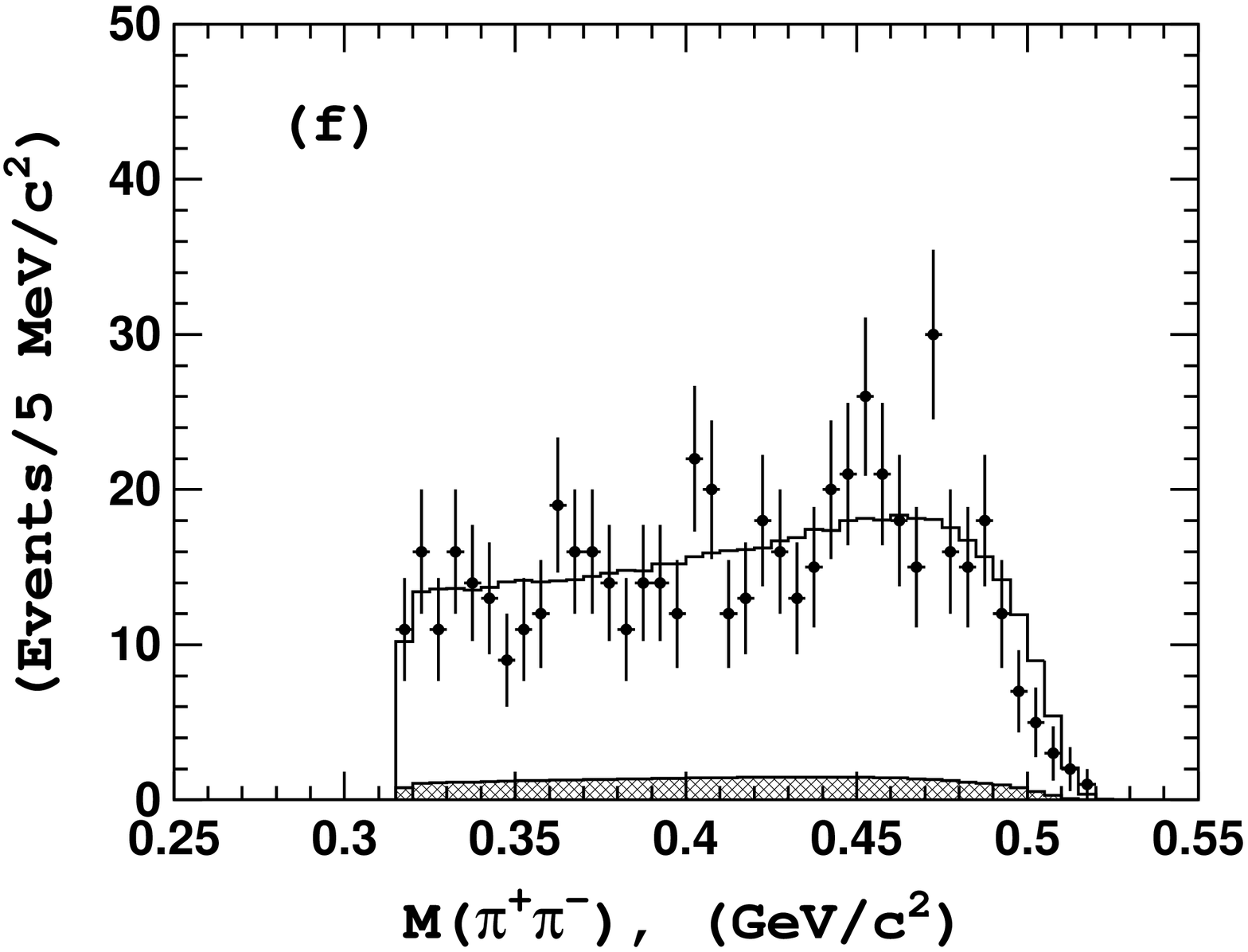} \\
  \caption{Comparison of fit results (open histogram) with
    experimental data (points with error bars) for events in the $\Uo$
    (a,b), $\Ut$ (c,d), and $\Uth$ (e,f) signal
    regions.  The hatched histogram shows the background component.}
\label{fig:y3spp-f-hh}
\end{figure}

Dalitz distributions of events in the $\Ut$ sidebands and signal regions
are shown in Figs.~\ref{fig:ynspp-b-dp}(a) and~\ref{fig:ynspp-b-dp}(b),
respectively, where $M(\Un\pi)_{\max}$ is the maximum invariant mass of
the two $\Un\pi$ combinations. This is used to combine $\Un\pi^+$ and
$\Un\pi^-$ events for visualization only.
Two horizontal bands are evident in the $\Ut\pi$ system near 
$112.6$~GeV$^2/c^4$ and $113.3$~GeV$^2/c^4$, where the distortion from 
straight lines is due to interference with other intermediate states,
as demonstrated below. One-dimensional invariant mass projections for 
events in the $\Un$ signal regions are shown in Fig.~\ref{fig:y3spp-f-hh},
where two peaks are observed in the $\Un\pi$ system near $10.61\,\gevm$
and $10.65\,\gevm$. In the following we refer to these structures as
$Z_b(10610)$ and $Z_b(10650)$, respectively. 

We parameterize the $\Uf\to\Un\pp$ three-body decay amplitude by:
\begin{equation}
M = A_{Z_1} + A_{Z_2} + A_{f_0} + A_{f_2} + A_{\rm nr},
\label{eq:model}
\end{equation}
where $A_{Z_1}$ and $A_{Z_2}$ are amplitudes to account for contributions
from the $Z_b(10610)$ and $Z_b(10650)$, respectively. Here we assume that the 
dominant contributions come from amplitudes that preserve the orientation
of the spin of the heavy quarkonium state and, thus, both pions in the
cascade decay $\Uf\to Z_b\pi\to\Un\pp$ are emitted in an $S$-wave with
respect to the heavy quarkonium system. As demonstrated in 
Ref.~\cite{FPCP-conf}, angular analyses support this assumption.
Consequently, we parameterize the observed $Z_b(10610)$ and $Z_b(10650)$
peaks with an $S$-wave Breit-Wigner function 
$BW(s,M,\Gamma)=\frac{\sqrt{M\Gamma}}{M^2-s-iM\Gamma}$,
where we do not consider possible $s$-dependence of the resonance
width. To account for the possibility of $\Uf$ decay to both
$Z_b^+\pi^-$ and $Z_b^-\pi^+$, the amplitudes $A_{Z_1}$ and $A_{Z_2}$
are symmetrized with respect to $\pi^+$ and $\pi^-$ transposition. 
Using isospin symmetry, the resulting amplitude is written as
\begin{equation}
A_{Z_k}=a_{Z_k}e^{i\delta_{Z_k}}(BW(s_1,M_k,\Gamma_k)+BW(s_2,M_k,\Gamma_k)),
\end{equation}
where $s_1 = M^2[\Un\pi^+]$, $s_2 = M^2[\Un\pi^-]$. The relative amplitudes
$a_{Z_k}$, phases $\delta_{Z_k}$, masses $M_k$ and widths $\Gamma_k$
($k = 1,2$) are free parameters. We also include the 
$A_{f_0}$ and $A_{f_2}$ amplitudes to account for possible contributions
in the $\pp$ channel from the $f_0(980)$ scalar and $f_2(1270)$ tensor states,
respectively. The inclusion of these two states is needed to describe the
shape of the $M(\pp)$ spectrum around and above $M(\pp)=1.0\,\gevm$ for
the $\Uo\pp$ final state (see Fig.~\ref{fig:y3spp-f-hh}).
We use a Breit-Wigner function to parameterize the
$f_2(1270)$ and a coupled-channel Breit-Wigner function~\cite{Flatte}
for the $f_0(980)$. The mass and width of the $f_2(1270)$ state are
fixed at their world average values~\cite{PDG}; the mass and the coupling
constants of the $f_0(980)$ state are fixed at values determined from the
analysis of  $B^+\to K^+\pp$: $M[f_0(980)]=950$~MeV/$c^2$, $g_{\pi\pi}=0.23$,
$g_{KK}=0.73$~\cite{kpp}.

Following suggestions in Ref.~\cite{Voloshin:2007dx}, 
the non-resonant amplitude $A_{\rm nr}$ is parameterized as 
$A_{\rm nr} = a^{\rm nr}_1\, e^{i\delta^{\rm nr}_1} +
              a^{\rm nr}_2\, e^{i\delta^{\rm nr}_2} \, s_3$,
where $s_3 = M^2(\pp)$ ($s_3$ is not an independent variable and can be
expressed via $s_1$ and $s_2$ but we use it here for clarity),
$a^{\rm nr}_1$, $a^{\rm nr}_2$, $\delta^{\rm nr}_1$ and $\delta^{\rm nr}_2$
are free parameters of the fit.

The logarithmic likelihood function ${\cal{L}}$ is then constructed as 
\begin{equation}
{\cal{L}} = -2\sum{\log(f_{\rm sig}S(s_1,s_2) + (1-f_{\rm sig})B(s_1,s_2))},
\end{equation}
where $S(s_1,s_2)$ is the density of signal events $|M(s_1,s_2)|^2$ 
convolved with the detector resolution function, $B(s_1,s_2)$ describes
the combinatorial background that is considered to be constant and
$f_{\rm sig}$ is the fraction of signal events in the data sample. 
Both $S(s_1,s_2)$ and $B(s_1,s_2)$ are efficiency corrected.

In the fit to the $\Uo\pp$ and $\Ut\pp$ samples, the amplitudes and phases
of all of the components are allowed to float. However, in the $\Uth\pp$
samples the available phase space is significantly smaller and contributions
from the $f_0(980)$ and $f_2(1270)$ channels are not well constrained.
Since the fit to the $\Uth\pp$ signal is insensitive to the presence of
these two components, we fix their amplitudes at zero. Due to the very
limited phase space available in the $\Uf\to\Uth\pp$ decay, there is a
significant overlap between the two processes $\Uf\to Z_b^+\pi^-$ and
$\Uf\to Z_b^-\pi^+$. 

Results of the fits to $\Uf\to\Un\pp$ signal events are shown in
Fig.~\ref{fig:y3spp-f-hh}, where one-dimensional projections of the
data and fits are compared. Numerical results are summarized in
Table~\ref{tab:results}, where the relative normalization is defined as
$a_{Z_2}/a_{Z_1}$ and the relative phase as $\delta_{Z_2}-\delta_{Z_1}$.
The combined statistical significance of the two peaks exceeds $10\,\sigma$
for all tested models and for all $\Un\pp$ channels.

The main source of systematic uncertainties in the analysis of $\Uf\to\Un\pp$
channels is due to uncertainties in the parameterization of the decay 
amplitude. We fit the data with modifications of the nominal model 
(described in Eq.~\ref{eq:model}). In particular, we vary the $M(\pp)$
dependence of the non-resonant amplitude $A_{\rm nr}$, include a $D$-wave
component into $A_{\rm nr}$, include the $f_0(600)$ state, etc. The
variations in the extracted $Z_b$ parameters determined from fits with
modified models are taken as estimates of the model uncertainties. Other
major sources of systematic error include variation of the reconstruction
efficiency over the Dalitz plot and uncertainty in the c.m.\ energy.
Systematic effects associated with uncertainties in the description of 
the combinatorial background are found to be negligible. The overall
systematic errors are quoted in Table~\ref{tab:results}.

\begin{figure}[!tb]
\vspace*{1mm}
\includegraphics[width=0.23\textwidth]{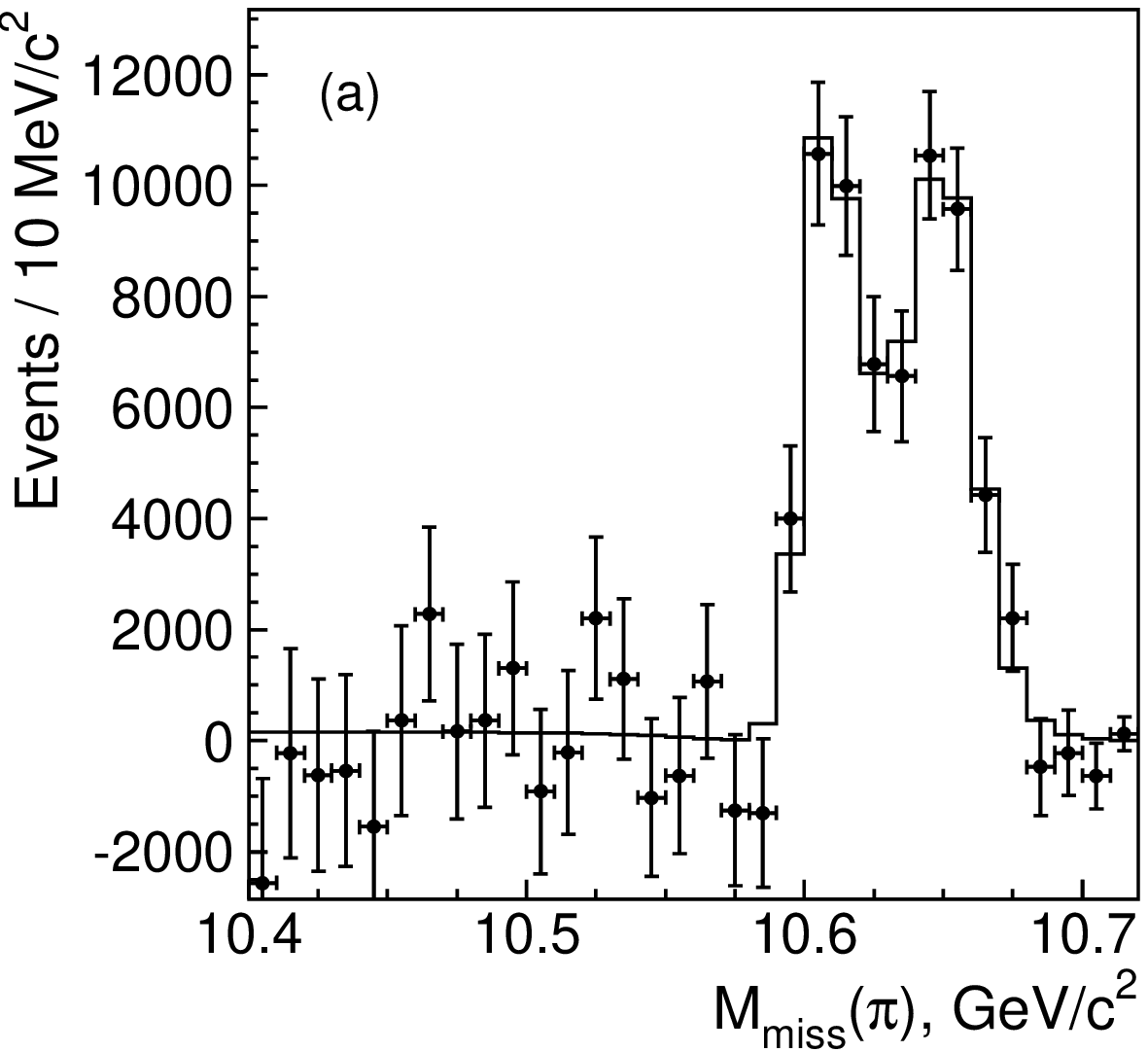} \hfill
\includegraphics[width=0.23\textwidth]{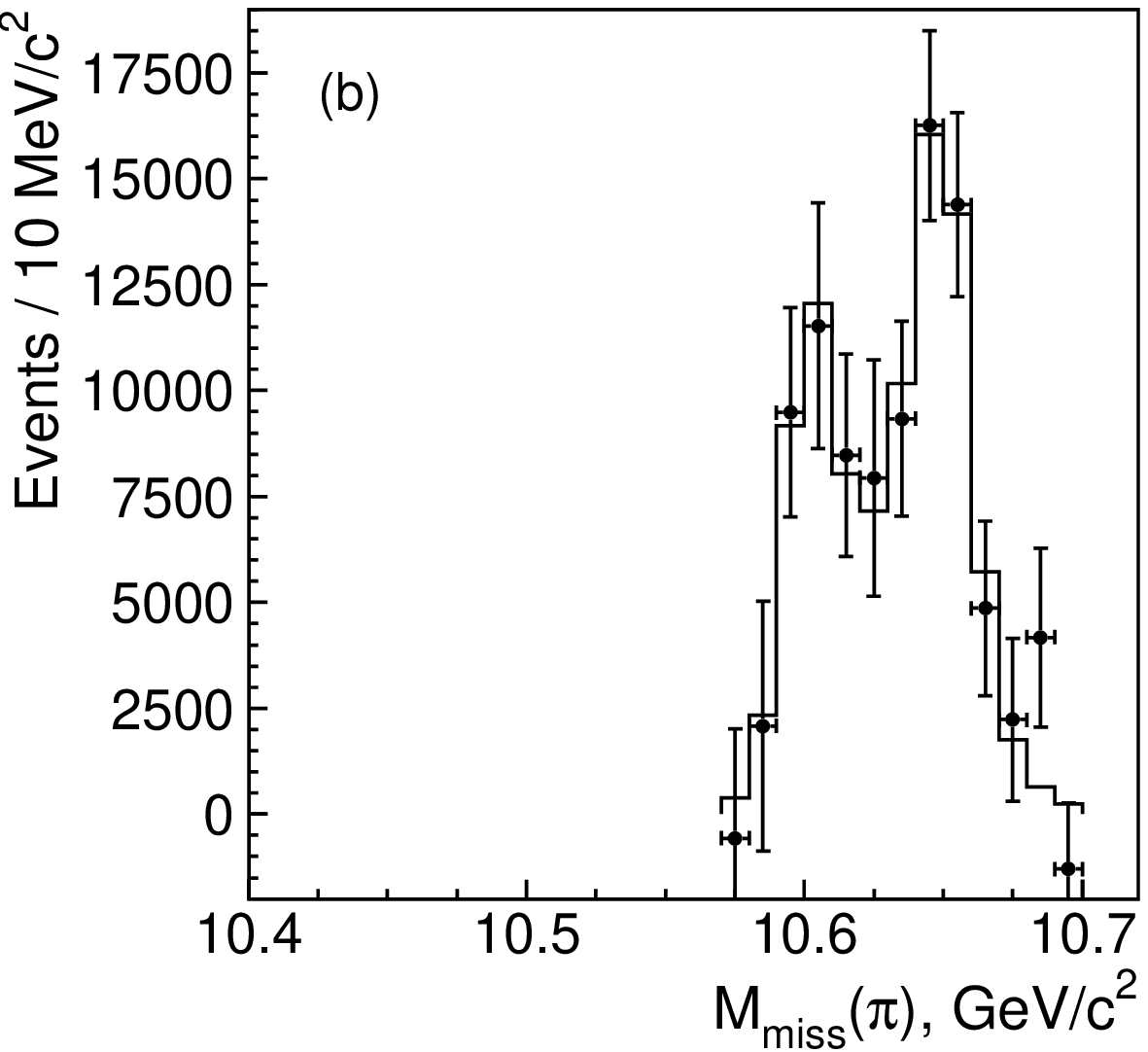}
\caption{The (a) $\hb$ and (b) $\hbp$ yields as a function of $\mmp$ 
(points with error bars) and results of the fit (histogram).}
\label{fig:mhbpi}
\end{figure}

To study the resonant substructure of the $\Uf\to\hbn\pp$ ($m=1,2$)
decays we measure their yield as a function of the $\hb\pipm$ invariant
mass. The decays are reconstructed inclusively using the missing mass of
the $\pp$ pair, $\mmpp$. We fit the $\mmpp$ spectra in bins of $\hb\pipm$
invariant mass, defined as the missing mass of the opposite sign pion,
$M_{\rm miss}(\pimp)$. We combine the $\mmpp$ spectra for the corresponding
$\mmpip$ and $\mmpim$ bins and we use half of the available $\mmp$ range
to avoid double counting.

\begin{table*}[!tb]
  \caption{Comparison of results on $Z_b(10610)$ and $Z_b(10650)$ parameters
           obtained from $\Uf\to\Un\pp$ ($n=1,2,3$) and $\Uf\to \hbn\pp$ 
           ($m=1,2$) analyses. }
  \medskip
  \label{tab:results}
\centering
  \begin{tabular}{lccccc} \hline \hline
Final state & $\Upsilon(1S)\pi^+\pi^-$      &
              $\Upsilon(2S)\pi^+\pi^-$      &
              $\Upsilon(3S)\pi^+\pi^-$      &
              $h_b(1P)\pi^+\pi^-$           &
              $h_b(2P)\pi^+\pi^-$
\\ \hline
           $M[Z_b(10610)]$, MeV/$c^2$       &
           $10611\pm4\pm3$                  &
           $10609\pm2\pm3$                  &
           $10608\pm2\pm3$                  &
           $10605\pm2^{+3}_{-1}$            &
           $10599{^{+6+5}_{-3-4}}$
 \\
           $\Gamma[Z_b(10610)]$, MeV  &
           $22.3\pm7.7^{+3.0}_{-4.0}$       &
           $24.2\pm3.1^{+2.0}_{-3.0}$       &
           $17.6\pm3.0\pm3.0$               &
           $11.4\,^{+4.5+2.1}_{-3.9-1.2}$   &
           $13\,^{+10+9}_{-8-7}$
 \\
           $M[Z_b(10650)]$, MeV/$c^2$       &
           $10657\pm6\pm3$                  &
           $10651\pm2\pm3$                  &
           $10652\pm1\pm2$                  &
           $10654\pm3\,{^{+1}_{-2}}$        &
           $10651{^{+2+3}_{-3-2}}$
 \\
           $\Gamma[Z_b(10650)]$, MeV  &
           $16.3\pm9.8^{+6.0}_{-2.0}$~      &
           $13.3\pm3.3^{+4.0}_{-3.0}$       &
           $8.4\pm2.0\pm2.0$                &  
           $20.9\,^{+5.4+2.1}_{-4.7-5.7}$   & 
           $19\pm7\,^{+11}_{-7}$ 
 \\
           Rel. normalization               &
           $0.57\pm0.21^{+0.19}_{-0.04}$    &
           $0.86\pm0.11^{+0.04}_{-0.10}$    &
           $0.96\pm0.14^{+0.08}_{-0.05}$    &
           $1.39\pm0.37^{+0.05}_{-0.15}$    &
           $1.6^{+0.6+0.4}_{-0.4-0.6}$
 \\
           Rel. phase, degrees              &
           $58\pm43^{+4}_{-9}$              &
           $-13\pm13^{+17}_{-8}$            &
           $-9\pm19^{+11}_{-26}$            &
           $187^{+44+3}_{-57-12}$           &
           $181^{+65+74}_{-105-109}$   
\\
\hline \hline
\end{tabular}
\end{table*}

Selection requirements and the $\mmpp$ fit procedure are described in
detail in Ref.~\cite{Belle_hb}. We consider all well reconstructed and
positively identified $\pp$ pairs in the event. Continuum $\ee\to
q\bar{q}$ ($q=u,\;d,\;s$) background is suppressed by a requirement on
the ratio of the second to zeroth Fox-Wolfram moments
$R_2<0.3$~\cite{Fox-Wolfram}.
The fit function is a sum of peaking components due to dipion
transitions and combinatorial background. The positions of all peaking
components are fixed to the values measured in
Ref.~\cite{Belle_hb}. In the case of the $\hb$ the peaking components
include signals from $\Uf\to\hb$ and $\Uf\to\Ut$ transitions,
and a reflection from the $\Uth\to\Uo$ transition, where the $\Uth$ is
produced inclusively or via initial state radiation. Since the
$\Uth\to\Uo$ reflection is not well constrained by the fits, we
determine its normalization relative to the $\Uf\to\Ut$ signal from
the exclusive $\uu\pp$ data for every $\mmp$ bin.
In case of the $\hbp$ we use a smaller $\mmpp$ range than in
Ref.~\cite{Belle_hb}, $\mmpp<10.34\,\gevm$, to exclude the region of
the $\ks\to\pp$ reflection. The peaking components include the
$\Uf\to\hbp$ signal and a $\Ut\to\Uo$ reflection. To constrain the
normalization of the $\Ut\to\Uo$ reflection we use exclusive $\uu\pp$
data normalized to the total yield of the reflection in the inclusive
data. Systematic uncertainty in the latter number is included in the
error propagation.
The combinatorial background is parameterized by a Chebyshev
polynomial. We use orders between 6 and 10 for the $\hb$ [the order
decreases monotonically with the $\mmp$] and orders between 6 and 8
for the $\hbp$.

The results for the yield of $\Uf\to\hbn\pp$ ($m=1,2$) decays as a
function of the $\mmp$ are shown in Fig.~\ref{fig:mhbpi}. The 
distribution for the $\hb$ exhibits a clear two-peak structure without a
significant non-resonant contribution. The distribution for the $\hbp$
is consistent with the above picture, though the available phase-space
is smaller and uncertainties are larger. We associate the two peaks with
the production of the $\zbo$ and $\zbt$. To fit the $\mmp$ distributions
we use the expression
\begin{equation}
|BW_1(s,M_1,\Gamma_1)+ae^{i\phi}BW_1(s,M_2,\Gamma_2)
+be^{i\psi}|^2\frac{qp}{\sqrt{s}}.
\label{hb_fit_fun}
\end{equation}
Here $\sqrt{s}\equiv\mmp$; the variables $M_k$, $\Gamma_k$ ($k=1,2$),
$a$, $\phi$, $b$ and $\psi$ are free parameters; $\frac{qp}{\sqrt{s}}$
is a phase-space factor, where $p$ ($q$) is the momentum of the pion
originating from the $\Uf$ ($\zb$) decay measured in the rest frame of
the corresponding mother particle.  The $P$-wave Breit-Wigner
amplitude is expressed as
$BW_1(s,M,\Gamma)=\frac{\sqrt{M\,\Gamma}\,F\,(q/q_0)}{M^2-s-iM\,\Gamma}.$
Here $F$ is the $P$-wave Blatt-Weisskopf form factor
$F=\sqrt{\frac{1+(q_0R)^2}{1+(qR)^2}}$~\cite{blatt-weisskopf}, $q_0$
is a daughter momentum calculated with pole mass of its mother,
$R=1.6\,\gev^{-1}$.  The function (Eq.~\ref{hb_fit_fun}) is convolved with
the detector resolution function ($\sigma=5.2\,\mevm$), integrated
over the $10\,\mevm$ histogram bin and corrected for the
reconstruction efficiency. The fit results are shown as solid
histograms in Fig.~\ref{fig:mhbpi} and are summarized in
Table~\ref{tab:results}. 
We find that the non-resonant contribution is consistent with zero
[significance is $0.3\,\sigma$ both for the $\hb$ and $\hbp$] in
accord with the expectation that it is suppressed due to heavy quark
spin-flip. In case of the $\hbp$ we improve the stability of the fit by
fixing the non-resonant amplitude to zero. The C.L. of the fit is 81\%
(61\%) for the $\hb$ [$\hbp$]. The default fit hypothesis is favored
over the phase-space fit hypothesis at the $18\,\sigma$
[$6.7\,\sigma$] level for the $\hb$ [$\hbp$].

To estimate the systematic uncertainty we vary the order of the Chebyshev
polynomial in the fits to the $\mmpp$ spectra; to study the effect of
finite $\mmp$ binning we shift the binning by half bin size; to study the
model uncertainty in the fits to the $\mmp$ distributions we remove [add]
the non-resonant contribution in the $\hb$ [$\hbp$] case; we increase the
width of the resolution function by 10\% to account for possible difference
between data and MC simulation. The maximum change of parameters for each
source is used as an estimate of its associated systematic error. 
We estimate an additional $1\,\mevm$ uncertainty in mass measurements
based on the difference between the observed $\Upsilon(nS)$ peak positions
and their world averages~\cite{Belle_hb}. The total systematic uncertainty
presented in Table~\ref{tab:results} is the sum in quadrature of 
contributions from all sources. The significance of the $\zbo$ and $\zbt$
including systematic uncertainties is $16.0\,\sigma$ [$5.6\,\sigma$] for
the $\hb$ [$\hbp$].

In conclusion, we have observed two charged bottomonium-like resonances, the
$\zbo$ and $\zbt$, with signals in five different decay channels, $\Un\pipm$
($n=1,2,3$) and $\hbn\pipm$ ($m=1,2$). The parameters of the resonances 
are given in Table~\ref{tab:results}. All channels yield consistent results.
Weighted averages over all five channels give
$M=10607.2\pm2.0\,\mevm$,
$\Gamma=18.4\pm2.4\,\mev$ for the $\zbo$ and
$M=10652.2\pm1.5\,\mevm$,
$\Gamma=11.5\pm2.2\,\mev$ for the $\zbt$, 
where statistical and systematic errors are added in quadrature. The $\zbo$
production rate is similar to that of the $\zbt$ for each of the five decay
channels. Their relative phase is consistent with zero for the final states
with the $\Un$ and consistent with 180 degrees for the final states with
$\hbn$. Production of the $Z_b$'s saturates the $\Uf\to\hbn\pp$ transitions
and accounts for the high inclusive $h_b(mS)$ production rate reported in
Ref.~\cite{Belle_hb}. Analyses of charged pion angular 
distributions~\cite{FPCP-conf} favor the $J^P=1^+$ spin-parity assignment
for both the $\zbo$ and $\zbt$. Since the $\Uf$ has negative $G$-parity, the
$\zb$ states have positive $G$-parity due to the emission of the pion.

The minimal quark content of the $\zbo$ and $\zbt$ is a four quark
combination. The measured masses of these new states are a few MeV/$c^2$
above the thresholds for the open beauty channels $B^*{\overline B}$
($10604.6$~MeV/$c^2$) and $B^* {\overline B^*}$ ($10650.2$~MeV/$c^2$).
This suggests a ``molecular'' nature of these new states, which might
explain most of their observed properties~\cite{milst}. The preliminary
announcement of these results triggered intensive discussion of other
possible interpretations~\cite{bugg,simonov,4q1,4q2}.

We are grateful to Alexander Milstein of BINP and Mikhail Voloshin of
University of Minnesota for fruitful discussions.
We thank the KEKB group for excellent operation of the accelerator,
the KEK cryogenics group for efficient solenoid operations, and the
KEK computer group and the NII for valuable computing and SINET4
network support.  We acknowledge support from MEXT, JSPS and Nagoya's
TLPRC (Japan); ARC and DIISR (Australia); NSFC (China); MSMT
(Czechia); DST (India); MEST, NRF, NSDC of KISTI, and WCU (Korea);
MNiSW (Poland); MES and RFAAE (Russia); ARRS (Slovenia); SNSF
(Switzerland); NSC and MOE (Taiwan); and DOE and NSF (USA).


\begin{thebibliography}{99}
\bibitem{foot1} In the text, for conciseness, we refer to the initial
state as the $\Uf$. However it is possible that the final states we
discuss have a source that is distinct from the $\Uf$; see  
K.-F.~Chen~{\etal} (Belle Collaboration), Phys.\ Pev.\ D {\bf 82},
091196R (2010).
\end{thebibliography}
\end{document}